\documentclass[aps,pra,amsmath,amssymb,showpacs,twocolumn,
preprintnumbers,floatfix,letterpaper]{revtex4}

\usepackage{epsfig}
\usepackage{graphicx}
\usepackage{graphics}
\usepackage{amsmath}
\usepackage{amssymb}
\usepackage{amsfonts}
\usepackage{color}
\usepackage{overpic}
\usepackage{rotating}

\newcommand{\be}{\begin{equation}}
\newcommand{\ee}{\end{equation}}
\newcommand{\ba}{\begin{eqnarray}}
\newcommand{\ea}{\end{eqnarray}}
\newcommand{\br}{\begin{array}}
\newcommand{\er}{\end{array}}

\newcommand{\A}{\boldsymbol A}
\newcommand{\x}{\boldsymbol x}
\newcommand{\z}{\boldsymbol z}

\renewcommand{\r}{\boldsymbol r}
\newcommand{\p}{\boldsymbol p}

\begin{document}
\bibliographystyle{apsrev}

\title{Displacement effect in strong-field atomic ionization by an
XUV pulse}

\author{Igor A. Ivanov}
\author{Anatoli S. Kheifets}
\author{Klaus Bartschat}

\altaffiliation{Permanent Address: Department of Physics and
Astronomy, Drake University, Des Moines, Iowa 50311, USA}

\affiliation{Research School of Physical Sciences, The Australian
National University, Australia Australian National University,
Canberra, ACT 0200, Australia}

\author{John Emmons}
\author{Sean M. Buczek}

\affiliation{Department of Physics and Astronomy, Drake University,
Des Moines, Iowa 50311, USA}

\author{Elena V. Gryzlova}
\author{Alexei N. Grum-Grzhimailo} 

\affiliation{Skobeltsyn Institute of Nuclear Physics, Lomonosov Moscow
State University, Moscow 119991, Russia}

\date{\today}

\begin{abstract}
We study strong-field atomic ionization driven by an XUV pulse with a
non\-zero displacement, the quantity defined as the integral of the
pulse vector potential taken over the pulse duration. We demonstrate
that the use of such pulses may lead to an extreme sensitivity of
the ionization process to subtle changes of the parameters of a driving
XUV pulse, in particular, the ramp-on/off profile and the carrier
envelope phase.  We illustrate this sensitivity for atomic
hydrogen and lithium driven by few-femto\-second XUV pulses with
intensity in the $\rm 10^{14}~W/cm^2$ range. We argue that the
observed effect is general and should modify strong-field ionization
of any atom, provided the ionization rate is sufficiently high.
\end{abstract}



\maketitle

Over the past decade, it has become possible to generate short and
intense pulses of coherent eXtreme \hbox{UltraViolet} (XUV) radiation.
Sub-femto\-second XUV pulses from high-order harmonic  generation (HHG)
sources \cite{Goulielmakis20062008,Sansone20102006} have been widely
used for time-resolved studies of atomic photo\-ionization in
attosecond streaking \cite{M.Schultze06252010} and interferometric
\cite{PhysRevLett.106.143002} experiments.  Few to tens of
femto\-second pulses from free-electron lasers (FEL)
\cite{ref2007,Shintake2008} have been instrumental for studying
complex dynamics governing both sequential and direct multiple
ionization processes \cite{0953-4075-46-16-164002}.

There are certain peculiarities of the photo\-ionization process in this
short-wavelength intense-field regime. A nonresonant radiation field
of high intensity can dress the single-electron continuum states, 
resulting in a distorted multi-peak structure of the photo\-electron
spectra \cite{PhysRevLett.108.253001}.  
The multi-peaked spectra are typically explained in
terms of the dressed-state picture \cite{0022-3700-13-6-018,Cohen04}, or by
dynamical interference in the emission process through the interplay
between the photo\-ionization and the AC Stark shift
\cite{PhysRevA.86.063412}.

In this Letter, we report yet another peculiarity of strong-field
atomic ionization. Under a certain condition, the photoionization
process becomes extremely sensitive to subtle changes of the driving
XUV pulse such as the ramp-on/off profile and the carrier envelope
phase (CEP). This condition can be formulated as a non\-zero net
displacement of the free electron, originally at rest, observed after
the end of the pulse. This displacement can be expressed as the
integral of the pulse vector potential calculated over the pulse
duration. [We assume that the vector potential is zero before and
after the pulse.]  For non\-zero displacement, we show that seemingly
insignificant changes of the pulse parameters may have a dramatic
effect on the photo\-electron spectrum and the photoelectron angular
distribution (PAD).

We explain this effect within the Kramers-Henneberger (KH) picture of
the ionization process, in which the so-called ``KH atom'' is moving in the
reference frame of the ionized electron.  The ionic potential seen by
the photo\-electron in this frame and averaged over its oscillations,
known as the KH potential, is distinctly different from the original
atomic potential but still capable of supporting infinitely many bound
states.  These bound states can be imaged by photo\-electron
spectroscopy and are responsible for unexpected stabilization of
atomic ionization by intense IR laser pulses \cite{Morales11102011}.
In the present case, a hardly noticeable change of the ramp on/off
profile  from linear to sine-squared of a long flat-top pulse results
in dramatically different KH potentials. This, in turn, alters
significantly the entire photo\-ionization process, thus resulting in
a strong variation of the photo\-electron spectrum as well as the
PAD.

To our knowledge, little attention has been paid to date to
strong-field ionization driven by the pulses with a non\-zero
displacement. About 20 years ago, the possibility of using such pulses
was discussed~\cite{Nefedov94}, but this work has never been followed
through.  In this Letter, we study ionization driven by such pulses
for realistic scenarios and suggest a specific recipe for possible
experimental tests.

We illustrate the ramp-on/off and CEP effects for hydrogen and
lithium atoms driven by $\sim$10 femto\-second pulses with peak
intensity in the $\rm 10^{14}~W/cm^2$ range.  Even though we use
specific XUV pulse parameters, the predicted effects appear to be
general and should modify strong-field ionization of any atom,
including resonant photo\-ionization, provided the ionization rate is
sufficiently high.  All examples presented in this Letter are for
linearly polarized electric field pulses along the $\hat{\z}$
direction, with the amplitude given by $E(t) = F(t) \sin(\omega t +
\delta)$, where $F(t)$ is the envelope function, $\omega$ is the
central frequency, and the CEP~$\delta$ is usually (except for one
case) chosen as~zero.

We describe the photo\-ionization process by the non\-relativistic
time-dependent Schr\"odinger equation (TDSE), which can be solved to a
very high degree of accuracy.  We restrict ourselves to the dipole
approximation, ignoring any non\-dipole, including magnetic field,
effects.  This is well justified for the chosen pulse parameters.  As
shown in~\cite{Kylstra00}, the degree of adiabaticity of the
laser-atom interaction does not modify significantly the breakdown of
the dipole approximation. Furthermore, the criterion $F_0/c \,
\omega^3 \ll 1$~\cite{Kylstra00} , where $F_0$ is the field amplitude
and $c$ is the speed of light, is very well fulfilled in our
calculations.  The latter condition corresponds to a displacement of
the electron due to the magnetic field by much less than the size of
the initial wave packet.

For the numerical treatment, we employed either the length or velocity
gauge of the electric dipole operator and three time-propagation
schemes (Crank-Nicolson~\cite{Crank-Nicolson}, matrix
iteration~\cite{PhysRevA.60.3125}, and short iterative
Lanczos~\cite{SIL}). All these schemes and gauges produced essentially
identical (within the thickness of the lines) results.  Exhaustive
tests were performed to ensure numerical stability with respect to the
space and time grids, as well as the number of partial waves coupled
in the solution of the TDSE.  In case of hydrogen, this stability and
accuracy were  used  to calibrate experimental laser
parameters such as the absolute intensity at the 1\%
level~\cite{PhysRevA.87.053411,Graydon2013}. In case of lithium, 
a very accurate theoretical description of the experimental
strong field ionization spectra was achieved \cite{PhysRevA.83.023413}.

\begin{figure}[t!]
\centering
\includegraphics[width=0.32\textwidth,clip=]{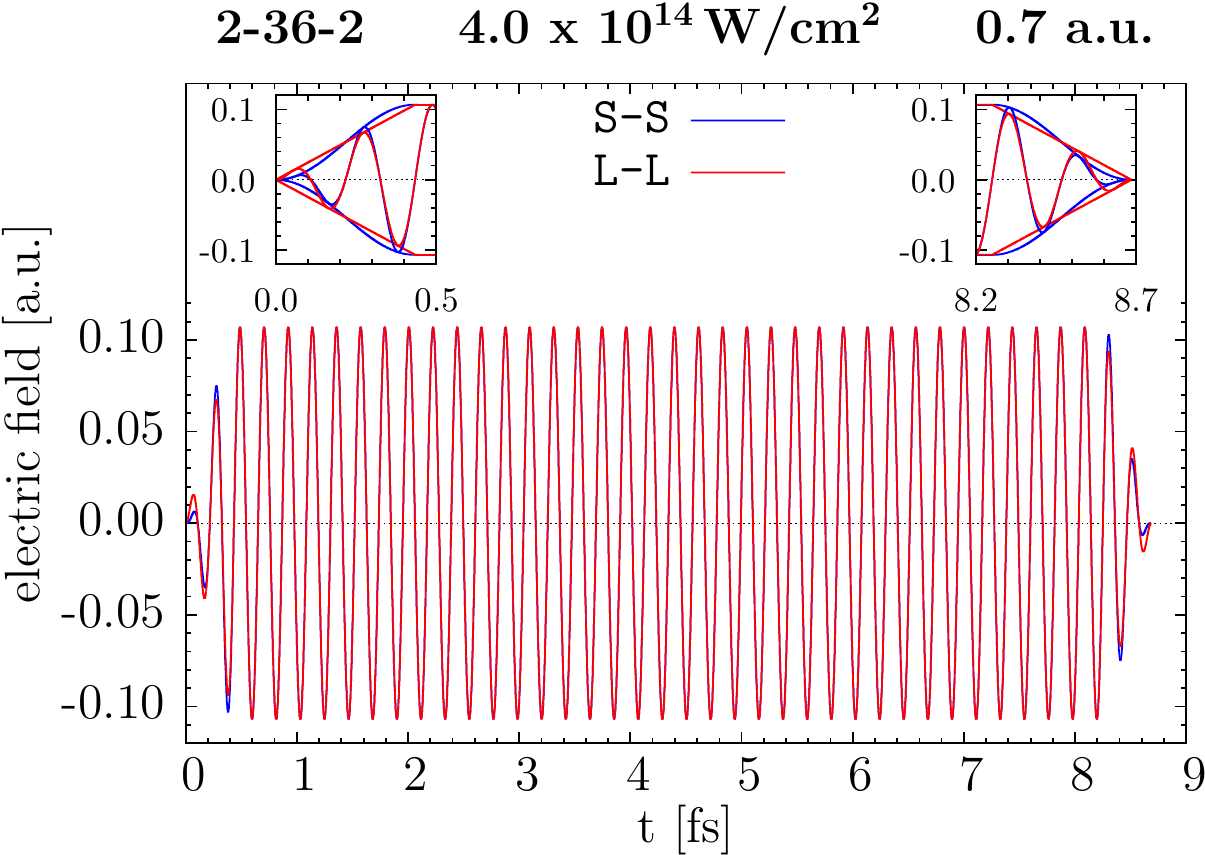} 

\vspace{2.0truemm}

\includegraphics[width=0.31\textwidth,clip=]{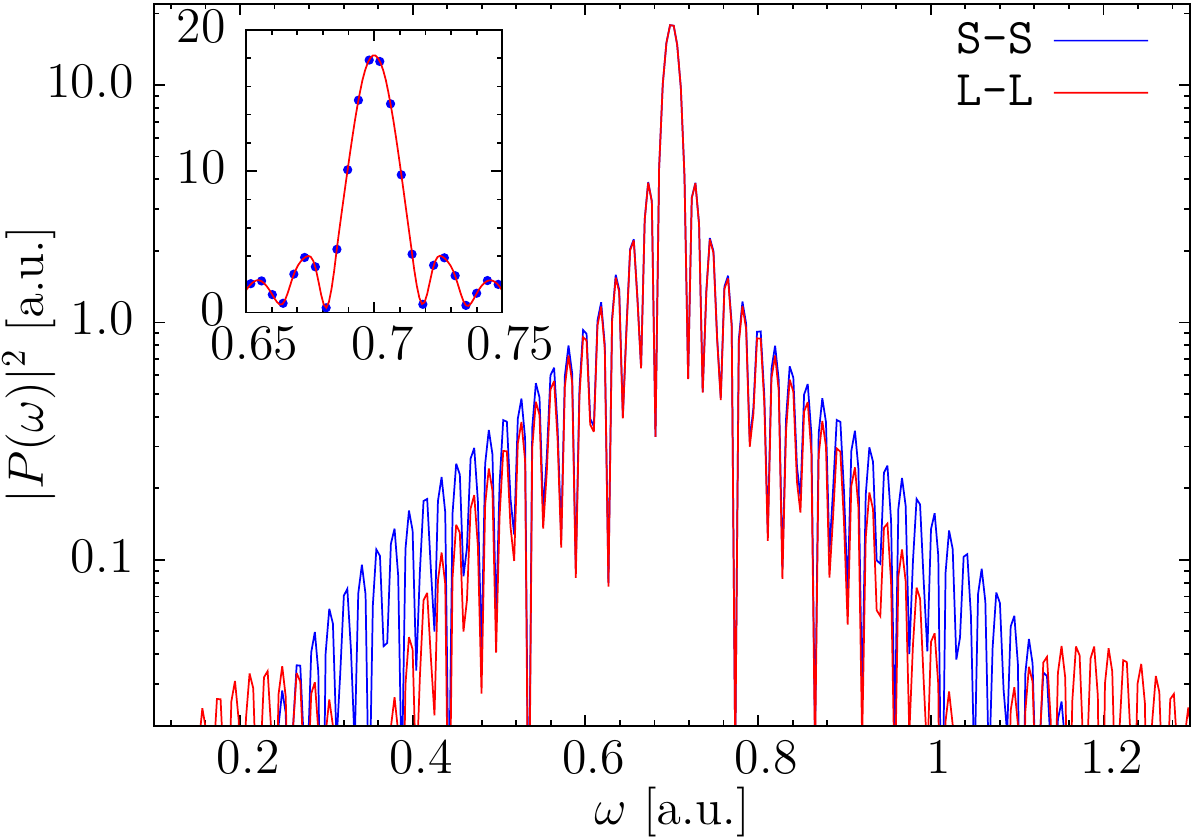} 

\caption{(Color online) Electric field (top) and Fourier spectrum
(bottom) of \hbox{2-36-2 L-L} and \hbox{S-S} pulses with central photon energy
0.7~a.u.~and peak intensity of $4.0 \times 10^{14}~$W/cm$^2$.  The
inserts in the top panel magnify the changes due to the small
differences in the ramp-on/off cycles. The pulses are identical in the plateau regime.
The \hbox{L-L} pulse in the bottom panel exhibits the higher side maxima.
The insert (solid circles are used for the S-S pulse to make it visible) shows that
the center of the frequency spectrum is virtually identical for the two cases.}
\label{fig:pulse}
\end{figure}

As a convenient numerical example, we consider the electric field
pulse with envelope functions of trapezoidal (linear ramp-on/off)
shape and sine-squared shape.  Both functions have the numerical
advantage that they start at true zero and can also be switched off
completely within a finite (not necessarily integer) number of cycles. In addition, an extended
plateau in the envelope function characterizes the amplitude
of the electric field.

Figure~\ref{fig:pulse} shows an example of two pulses, which we will
denote by ``2-36-2 S-S'' and ``2-36-2 L-L'', respectively.  Here
``$n_1$-$n_2$-$n_3$'' refers to the number of cycles in the ramp-on
($n_1$), the plateau ($n_2$), and the ramp-off ($n_3$), while ``S''
and ``L'' label sine-squared~(S) or linear~(L) ramp-on/off.  In this
particular example, the peak intensity is $4.0 \times
10^{14}~$W/cm$^2$, corresponding to a peak electric field amplitude of
0.107 atomic units (a.u.).  The central photon energy is 19~eV
(0.7~a.u.).  A similar pulse was studied recently in the context of
testing numerical
approaches~\cite{PhysRevA.83.013401,PhysRevA.88.055401}, except that
the central photon frequency was chosen to coincide with the
non\-relativistic \hbox{$1s$-$2p$} resonance transition energy in what
is expected to be predominantly a two-photon process.  We chose a
non\-resonant frequency significantly larger than the field-free
ionization potential in the present work, in order to avoid the
impression that the effects discussed below are limited to particular
resonant cases.

\begin{figure}[t!]
\centering
\includegraphics[width=0.32\textwidth,clip=]{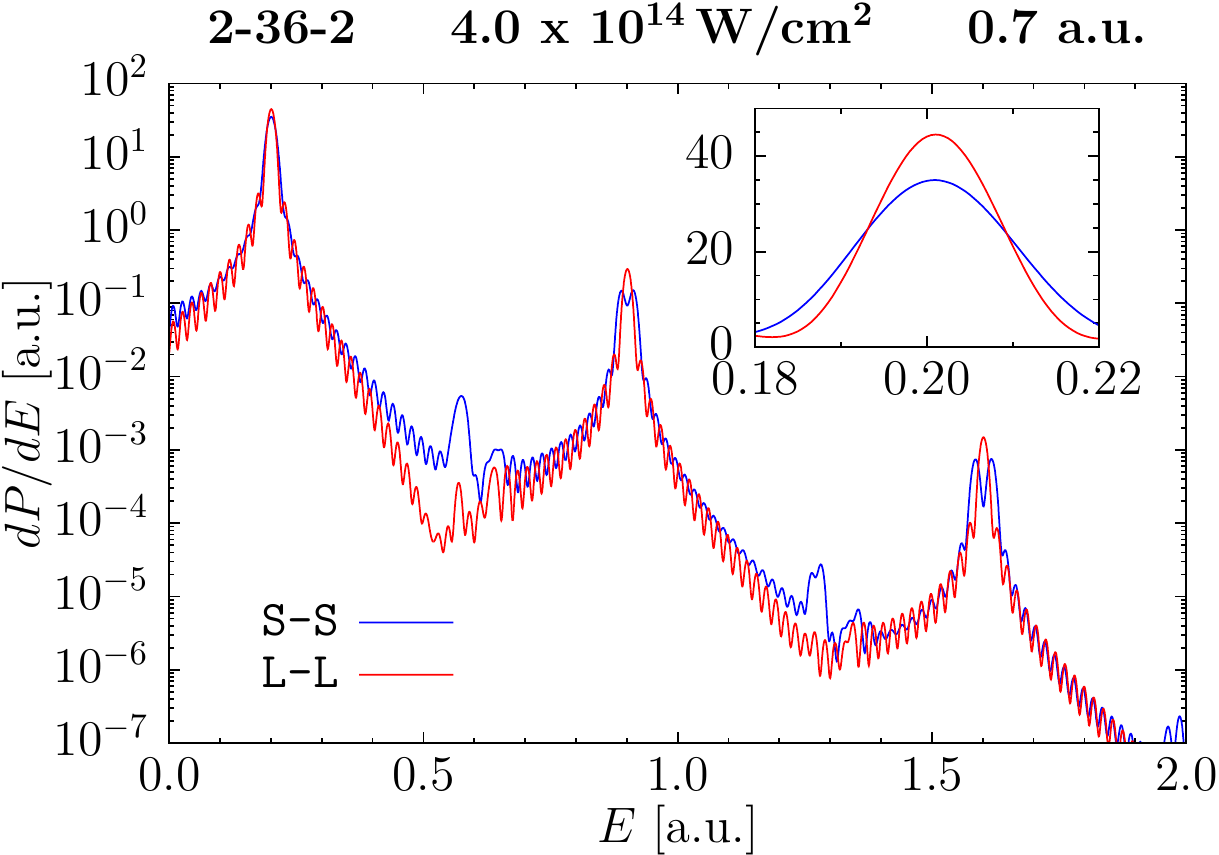} 

\vspace{2.0truemm}

\includegraphics[width=0.32\textwidth,clip=]{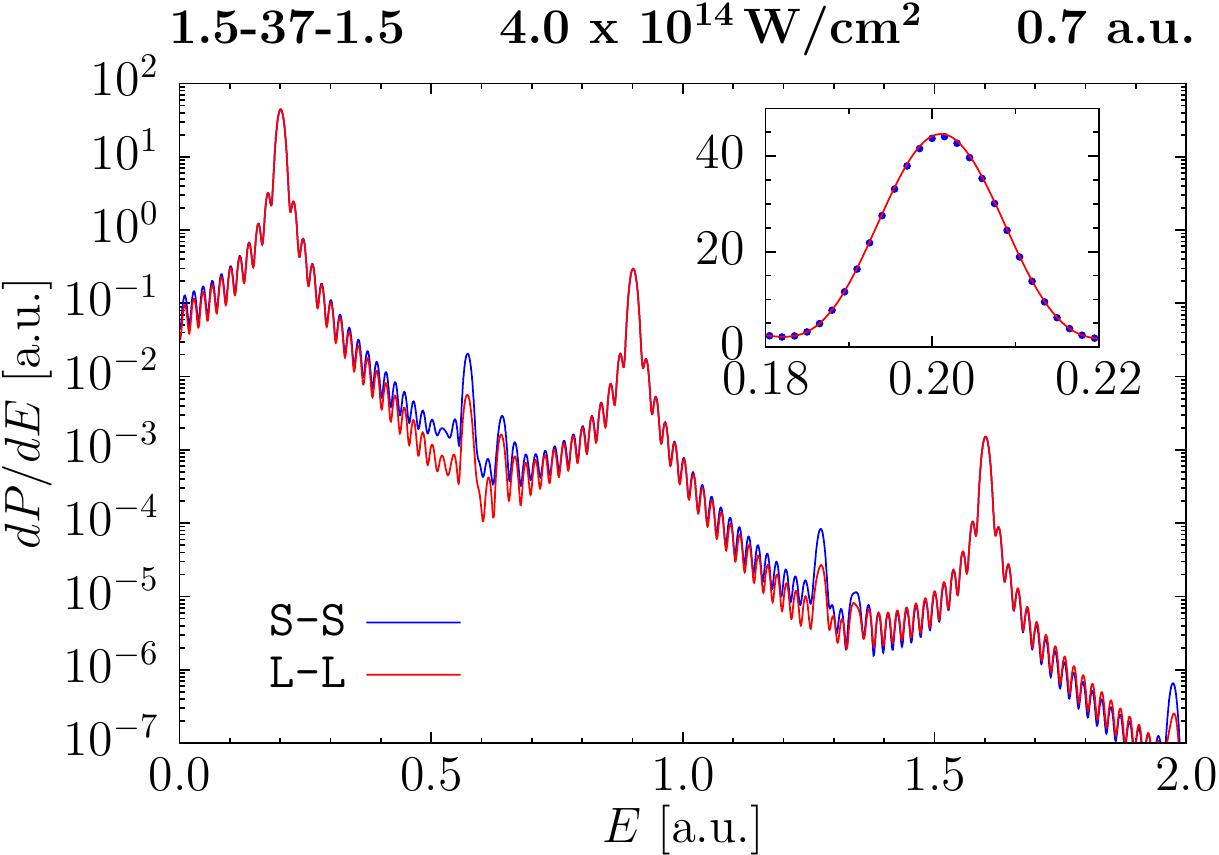} 

\caption{(Color online) Ejected-electron energy spectrum for
\hbox{2-36-2} and \hbox{1.5-37-1.5} pulses with central photon energy
0.7~a.u.~and peak intensity of $4.0 \times 10^{14}~$W/cm$^2$. For visibility, 
dots were used for the S-S results in the lower insert.}
\label{fig:spectrum}
\end{figure}

While the well-known multi-photon character in the ejected-electron
energy spectrum displayed on a logarithmic scale in
Fig.~\ref{fig:spectrum} may not look peculiar at all, the {\it insets}
show that the ramp-on/off effect can be substantial.  Not only does it
depend on the small difference in how the pulse is switched on and off
within a given number of optical cycles (o.c.), but also on how many
cycles are taken for the on/off steps.  Specifically, the dominant
single-photon peak displayed in the insets changes its height and
width when comparing the two \hbox{2-36-2} pulses, while virtually no
difference occurs for \hbox{1.5-37-1.5}.  Other peaks at higher
photo\-electron energies, corresponding to absorption of two and three
photons, are split into doublets.  Space does not allow for more
examples here, which we refer to future publications.  These results
may seem surprising, as both the \hbox{L-L} and \hbox{S-S} pulses have
very similar spectral content (cf.~bottom panel of
Fig.~\ref{fig:pulse}) and the phase of the Fourier decomposition (not
shown).

\begin{figure}[t!]
\centering
\includegraphics[width=0.23\textwidth,clip=]{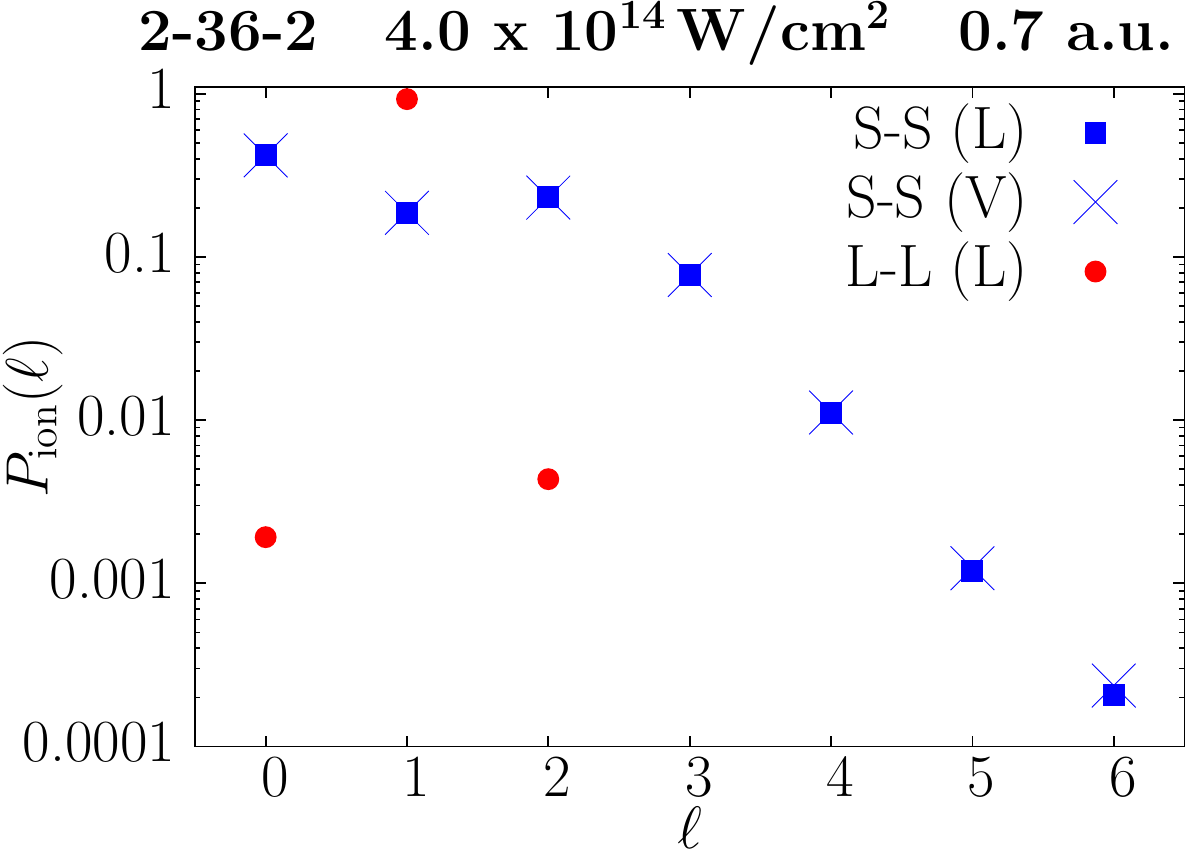}
\includegraphics[width=0.23\textwidth,clip=]{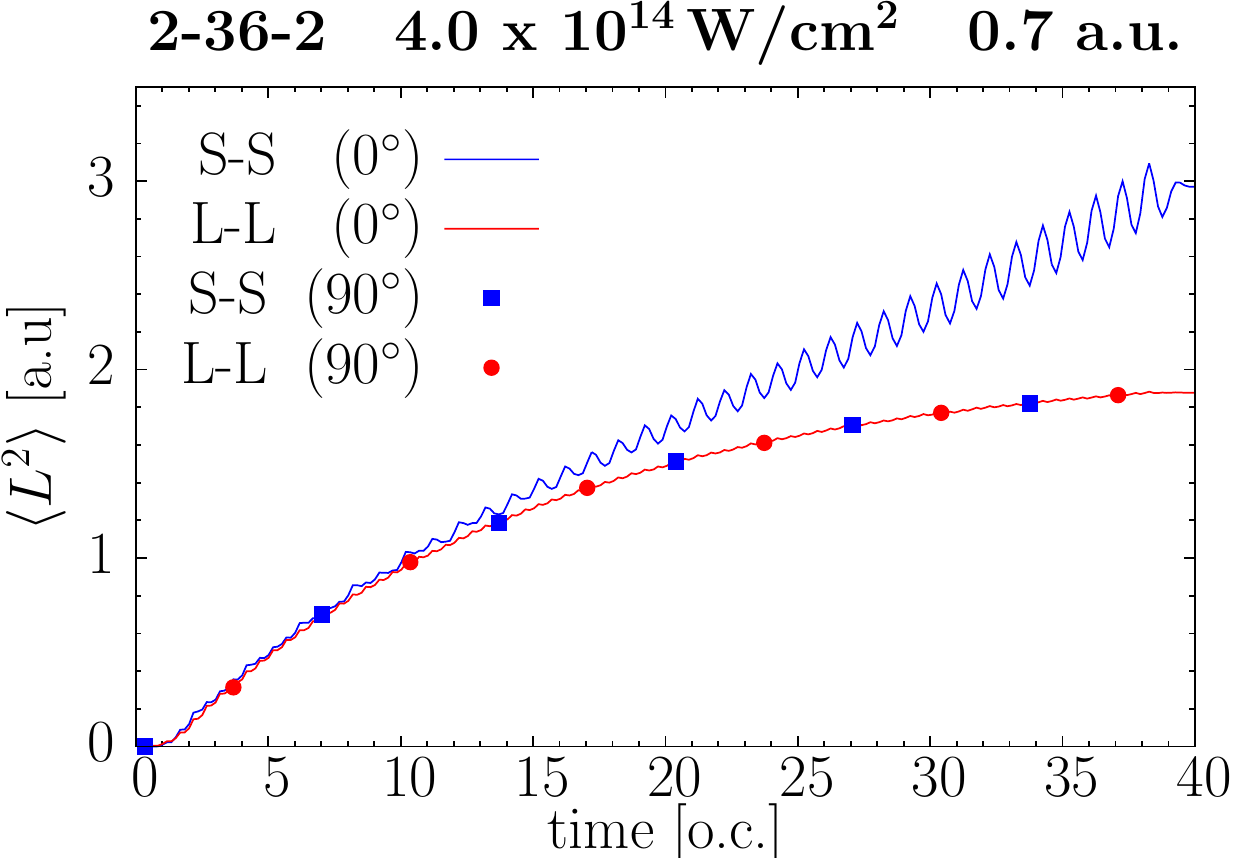}
\caption{(Color online) Left: Angular-momentum composition of the
ejected electron wave function after exposure to a \hbox{2-36-2} pulse
with central photon energy 0.7~a.u.~and peak intensity of $4.0 \times
10^{14}~$W/cm$^2$.  Note the broad distribution for the \hbox{2-36-2
S-S} pulse and the excellent agreement between the numerical
predictions obtained by independent computer codes in the length and
velocity gauges.
Right: Quantum mechanical expectation value of $\langle L^2 \rangle$,
as a function of time, for CPEs of $0^\circ$ and $90^\circ$.
}
\label{fig:L-dist}
\end{figure}

Further analysis revealed that not only the angle-integrated spectra
are very sensitive to the ramp-on/off.  The partial-wave decomposition
of the ionization probability, for example, and the evolution of the
expectation value $\langle L^2 \rangle$ as function of time, are
completely different for \hbox{2-36-2 S-S} and \hbox{2-36-2 L-L}
pulses.  This is shown in Fig.~\ref{fig:L-dist}.
While the $\langle L^2 \rangle$ expectation value in the presence of
the laser pulse is not a directly observable quantity (it is not gauge
invariant, but the right panel of Fig.~\ref{fig:L-dist} illustrates its evolution
if the velocity gauge is employed), the marked difference in its
behavior for \hbox{2-36-2 S-S} and \hbox{2-36-2 L-L} suggests that the
quantum evolution of the system proceeds very differently in these two
cases.  We also see that changing the CEP of the \hbox{S-S} pulse can
change the picture substantially. In fact, a CEP of 90$^\circ$ makes
the \hbox{2-36-2 S-S} pulse look ``normal'' again.

The partial-wave ($\ell$) decomposition of the ionization probability
(cf.~Fig.~\ref{fig:L-dist}), when computed after the end of the pulse,
is another gauge-invariant parameter that can be used to check the
partial-wave convergence of a calculation. In practice, the related
PAD is measured experimentally, but we first look at the
$\ell$-decomposition.

\begin{figure}[t!]
\centering
\includegraphics[width=0.33\textwidth,clip=]{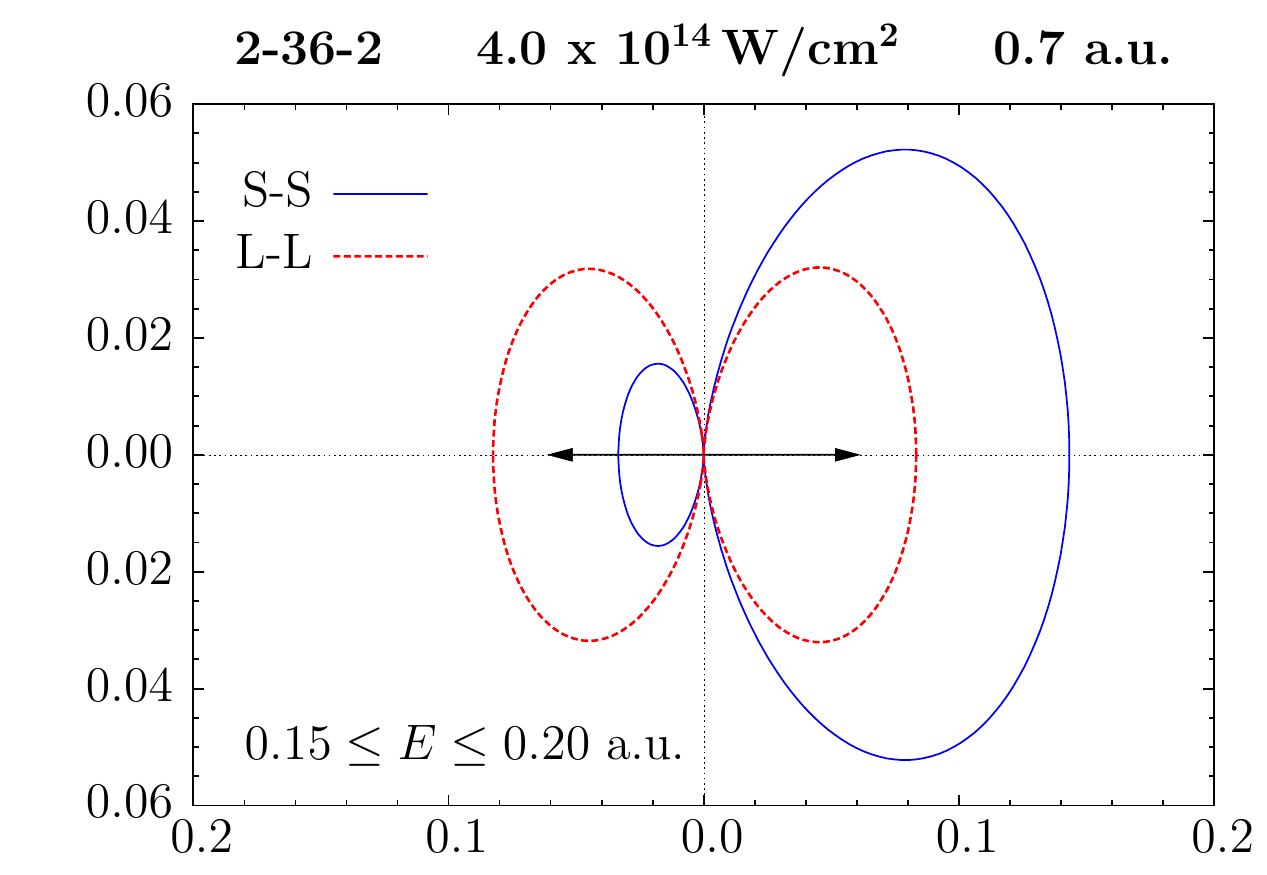} 

\caption{(Color online) PADs for the 
\hbox{2-36-2 S-S} (blue solid line) and \hbox{2-36-2 L-L} (red dashed line) pulse,
integrated over the energy interval \hbox{0.15-0.2~a.u.}
The arrow indicates the direction of the laser polarization axis. 
}
\label{fig:angdist}
\end{figure}

While the distribution is sharply peaked at $\ell=1$ for the
\hbox{2-36-2 L-L} pulse, as one would expect for a one-photon process,
Fig.~\ref{fig:L-dist} shows that it is broadly spread out for the
\hbox{2-36-2 S-S} pulse. As demonstrated in Fig.~\ref{fig:angdist},
the effect is, indeed, {\it observable} if the PAD is measured with an
{\it asymmetric energy window} around the central peak. Such windows
are typically set in experiments with reaction
microscopes~\cite{Ullrich2003}.  As seen in Fig.~\ref{fig:angdist},
the PADs obtained by integrating differential angle- and
energy-resolved ionization probabilities over the energy interval
\hbox{0.15~a.u.~$\le E \le 0.2$~a.u.} differ dramatically.

\begin{figure}[t!]
\centering
\includegraphics[width=0.24\textwidth,clip=]{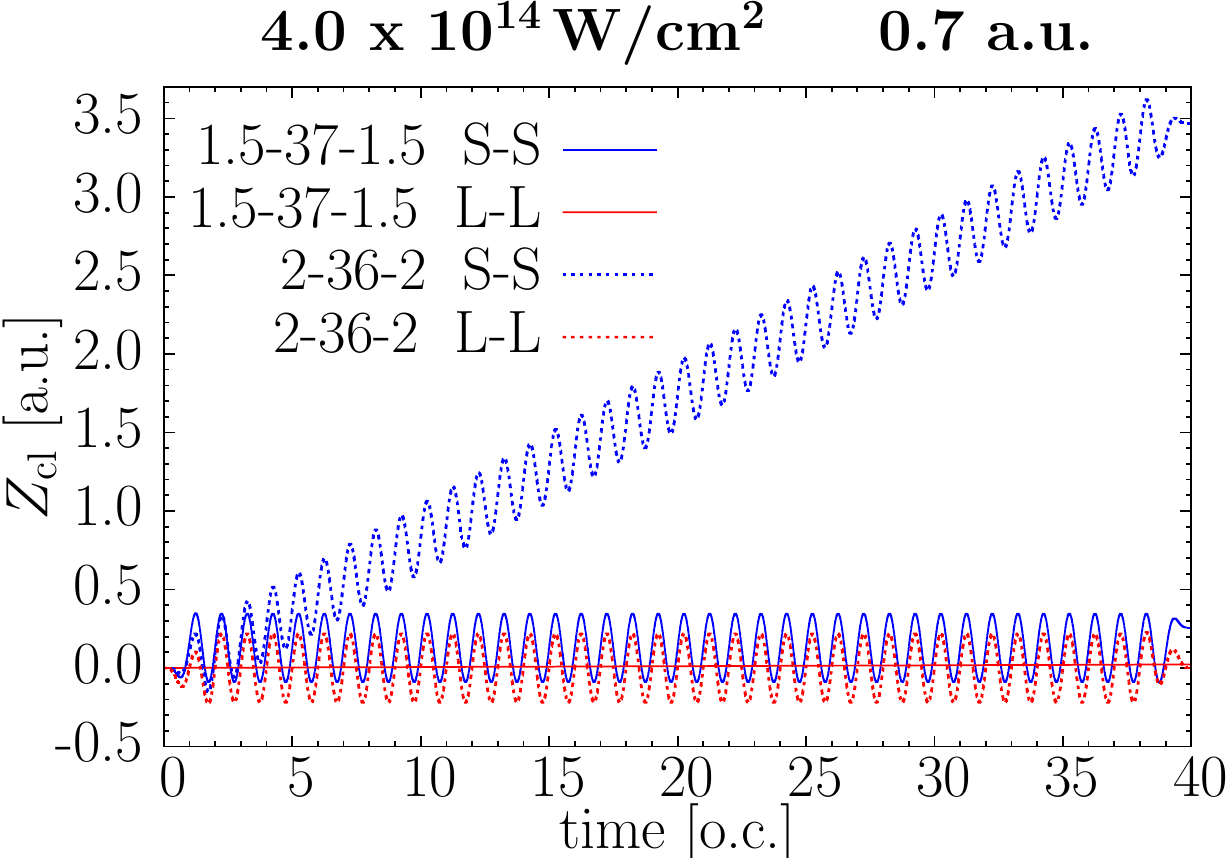} 
\includegraphics[width=0.23\textwidth,clip=]{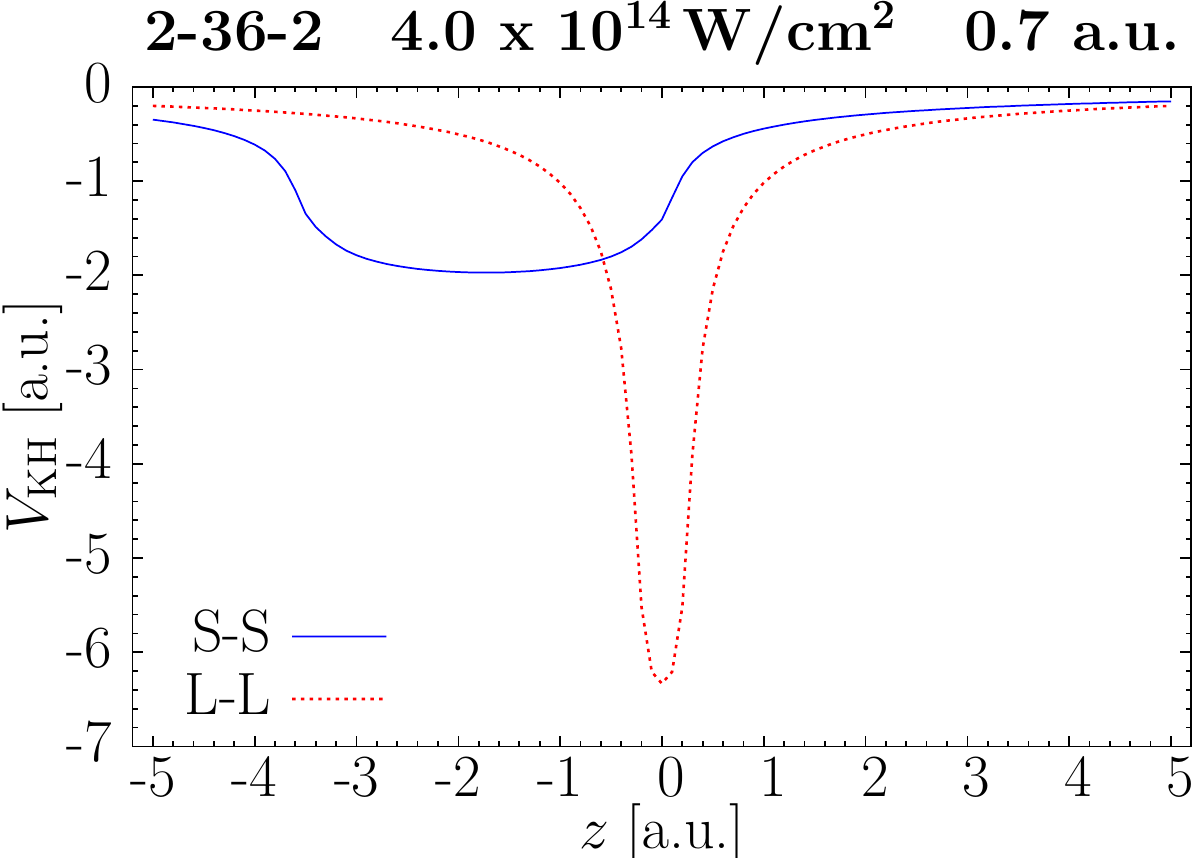}
\caption{(Color online) Left: Classical trajectory without Coulomb
field for an electron starting at the origin with zero speed under the
influence of the laser field for \hbox{1.5-37-1.5} and \hbox{2-36-2}
pulses with central photon energy 0.7~a.u.~and peak intensity of $4.0
\times 10^{14}~$W/cm$^2$.  
Right: Kramers-Henneberger potential along the
line running at the distance $\rho=0.1$ a.u.~parallel to the
polarization axis of the laser pulse for the \hbox{2-36-2 S-S} (blue)
solid and \hbox{2-36-2 L-L} (red) dashed lines.
}
\label{fig:classical}
\end{figure}

To explain these findings, we resort to the KH
picture of the ionization process \cite{kh1,PhysRevLett.21.838}. The
Hamiltonian operators in the KH gauge and the velocity gauge are
related by a canonical transformation generated by the operator
$
\hat T=\int_0^t \, \A(\tau) \cdot \hat\p \ d\tau ,\
$
where $\A(\tau)$ is the vector potential. This transformation yields
the Hamiltonian in the KH gauge:
\begin{equation}
\hat H_{\rm KH}= 
e^{{\rm i}\hat T}\hat H_{\rm V}e^{-{\rm i}\hat T}-{\partial \hat T\over \partial t}=
{\hat\p^2\over 2}+V(\r+ \x(t)) \, ,
\label{eq:kh}
\end{equation}
where $ \x(t)=\int_0^t {\A}(\tau)\, d\tau$, and $V(\r)$ is the
potential energy in the atomic field-free Hamiltonian displaced by
$\x(t)$, which is determined by the classical trajectory launched with
initial zero coordinate and velocity in a linearly polarized laser
field along the $\hat{\z}$ direction. For this geometry $\x(t)=Z_{\rm
cl}(t)\hat {\x}$. The quantity $ Z_{\rm cl}(t)$ is exhibited on the
left panel of Fig.~\ref{fig:classical} for various pulses. We see that
it is very different for the \hbox{2-36-2 S-S} pulse compared to the
\hbox{2-36-2 L-L} pulse or either one of the \hbox{1.5-37-1.5} pulses.

Different behaviour of $\x(t)$ leads to different Hamiltonians in the
KH picture.  This difference can be illustrated by
the so-called KH potential defined as
\begin{equation}
V_{KH}(\r)={1\over T_1} \int\limits_0^{T_1} V(\r+ \x(t))\, dt ,
\label{eq:kh2}
\end{equation}
where $T_1$ is the total pulse duration.  The KH potential in
Eq.~(\ref{eq:kh2}) is the zero-order term in the Fourier expansion of
the potential $V(\r+\x(t))$.  In many cases this term alone provides
enough information to understand qualitatively the effect of the laser
field on a system~\cite{Morales11102011}.  If necessary, corrections
to this simplified description can be generated by adding higher-order
terms of the Fourier expansion. We show the KH potentials on the right
panel of Fig.~\ref{fig:classical} for the \hbox{2-36-2 S-S} and
\hbox{2-36-2 L-L} pulses. Note that the KH potential for the
\hbox{2-36-2 L-L} pulse is nearly Coulombic, whereas for the
\hbox{2-36-2 S-S} case it is strongly distorted and far away from a
spherically symmetric form. This provides another explanation why the
angular-momentum distributions presented above for the \hbox{2-36-2
S-S} case are so broad.

\begin{figure}[h!]
\centering
\includegraphics[width=0.32\textwidth,clip=]{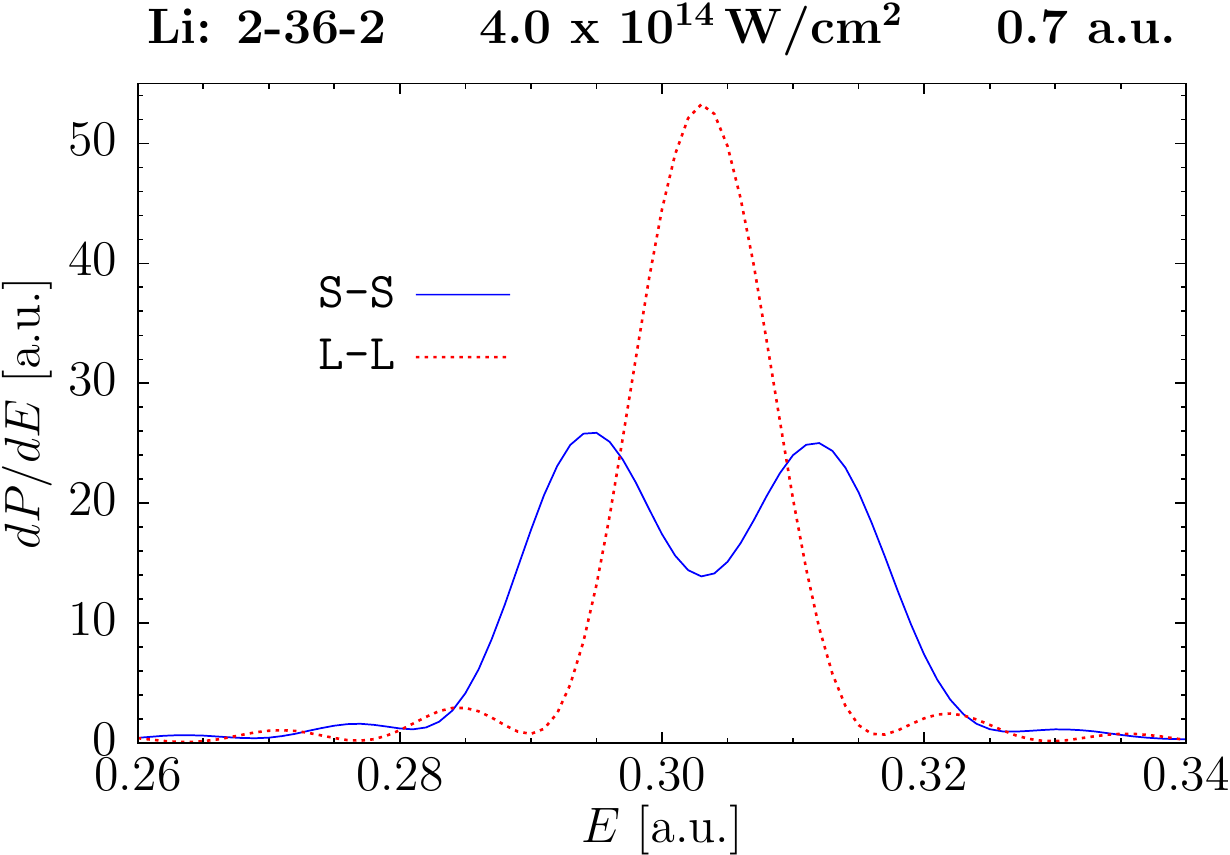}
\caption{(Color online) 
Ejected electron spectrum for Li for ionization by 
the \hbox{2-36-2 S-S} (blue) solid
and \hbox{2-36-2 L-L} (red) dashed lines 
pulses with central photon energy
0.5~a.u.~and peak intensity of $4.0 \times 10^{14}~$W/cm$^2$.}
\label{fig:lis}
\end{figure}

\begin{figure}[h!]
\centering
\includegraphics[width=0.32\textwidth,clip=]{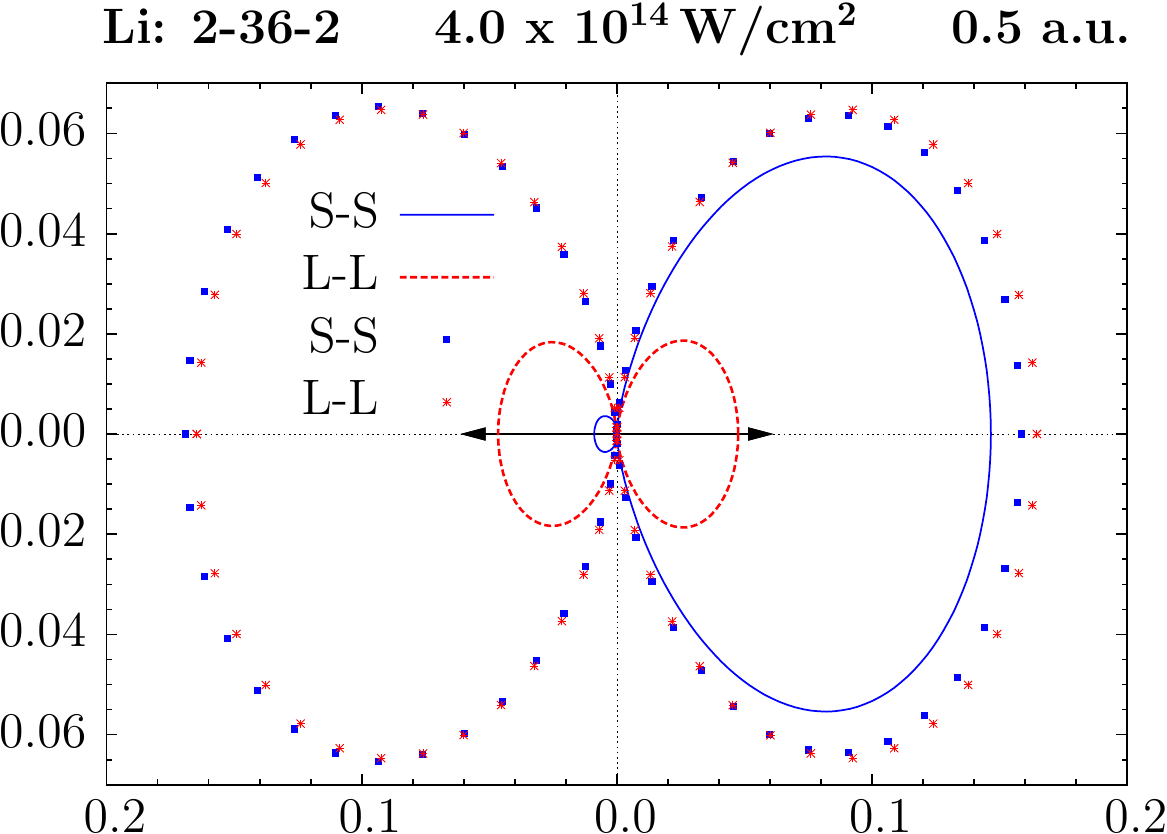}
\caption{(Color online) PADs for Li by pulses with central photon
energy 0.5~a.u.~and peak intensity of $4.0 \times 10^{14}~$W/cm$^2$.
Lines are for an asymmetric energy window, 0.25~a.u.$\;\le E
\le$0.30~a.u., while symbols are for a symmetric energy window,
0.25~a.u.$\;\le E \le$0.35~a.u., around the central peak.  The arrow
indicates the direction of the laser polarization axis.  }
\label{fig:Li-PAD}
\end{figure}

Because of its universal nature, this effect should be observable in any atom
and not just be restricted to the hydrogen case chosen for
illustration. Indeed, Fig.~\ref{fig:lis} displays ionization spectra for the
lithium atom driven by a similar set of S-S and L-L pulses with a
central frequency of 13.6~eV and peak intensity of $4.0
\times 10^{14}~$W/cm$^2$. The ramp on/off effect in the energy spectra
is very similar to that observed for the hydrogen atom. It also
manifests itself in the PADs integrated over the energy interval covering
approximately half of the ionization peak in Fig.~\ref{fig:lis}, while it
essentially disappears if a symmetric energy window is used.  
The latter is illustrated in Fig.~\ref{fig:Li-PAD}.

\vspace{2.0truemm} To summarize, we have demonstrated a significant,
and so far unexplored {\it for realistic scenarios}, effect of the
laser pulse ramp-on/off and CEP on atomic ionization in the strong
field regime when the driving  XUV pulse has a non\-zero displacement.
We attribute this effect to small changes in the initial conditions
launching significantly different classical electron trajectories.
This, in turn, leads to different Kramers-Henneberger potentials
experienced by the receding photo\-electron and results in
significantly different photo\-electron spectra, angular-momentum
compositions, and PADs.  

We illustrated the proposed effect using specific pulse parameters
that are not far from those presently available from HHG and FEL
sources.  Once we find a combination of the pulse parameters
describing the ramp-on/off and the CEP such that the displacement has
a non\-zero value, we may expect a dramatic effect in the energy
spectra and PADs. The stronger the field and the longer the
pulse, the more visible the effect should generally be.
Also, the ramp-on/off effect is very visible in resonant
photo\-ionization. We observed a strong modification of the
Autler-Townes doublet in hydrogen at the resonant photon energy of 3/8~a.u.
Details will be discussed in future publications.

An important issue, of course, concerns the occurrence of pulses
with a non\-zero displacement experimentally. 
To our knowledge, the existence of such pulses does not contradict Maxwell's
equations, nor any other known physical law.  
\citet{0953-4075-40-12-005} strongly favored pulses with zero 
displacement, in order to prevent the electron from leaving the laser interaction
region too early.  In practice, however, a displacement of a few atomic units
(c.f.~Fig.~\ref{fig:classical}) should not be unrealistic in light of the 
typical size of the laser focus.  

In fact, the requirement that the
net displacement is zero, i.e., that the integral of the vector
potential over the pulse duration vanishes, is very
restrictive. This constraint connects the pulse shape and its CEP, i.e., 
we cannot freely change one without changing the
other if we want to limit the pulse to cause zero displacement. We are
not aware of this restriction having ever been considered, for
example, in the design or interpretation of experiments on quantum
control. Theoretically at least, these parameters are varied
independently.

\thanks

The authors benefited greatly from many stimulating discussions with
Misha Ivanov, Igor Litvinyuk, and Peter Hannaford. One of the authors
(KB) wishes to thank the Australian National University for
hospitality.  This work was supported by the Australian Research
Council under Grant No.~DP120101805 (IAI and ASK), by the United
States National Science Foundation under Grants No.~PHY-1068140,
PHY-1430245, and the XSEDE allocation PHY-090031 (KB, JE, SMB), and by
the Russian Foundation for Basic Research under Grant No.~12-02-01123
(EG and ANG).  Resources of the Australian National Computational
Infrastructure (NCI) Facility were also employed.


\end{document}